\documentclass[journal=jacsat,manuscript=article]{achemso}
\usepackage{chemformula} 
\usepackage[T1]{fontenc} 
\usepackage{xcolor}
\usepackage{placeins}
\usepackage{amsmath}
\usepackage{graphicx}
\usepackage{booktabs}
\usepackage{float}
\usepackage{fancyhdr}
\usepackage{color}
\usepackage{tabularray}
\usepackage{array}
\usepackage{multirow}
\usepackage{ragged2e}
\usepackage[version=4]{mhchem} 
\usepackage{siunitx}    
\usepackage{geometry}
\author{Qianqian Ma}
\affiliation{Resource Geophysics Academy, Imperial College London, London, SW7 2BP, United Kingdom}
\alsoaffiliation{Department of Earth Science and Engineering, Imperial College London, London, SW7 2AZ, United Kingdom}
\author{Rukuan Chai}
\affiliation[Imperial College London]
{Department of Earth Science and Engineering, Imperial College London, London, SW7 2AZ, United Kingdom}
\author{Sajjad Foroughi}
\affiliation[Imperial College London]
{Department of Earth Science and Engineering, Imperial College London, London, SW7 2AZ, United Kingdom}
\author{Yanghua Wang}
\affiliation{Resource Geophysics Academy, Imperial College London, London, SW7 2BP, United Kingdom}
\alsoaffiliation{Department of Earth Science and Engineering, Imperial College London, London, SW7 2AZ, United Kingdom}
\author{Martin J. Blunt}
\affiliation[Imperial College London]
{Department of Earth Science and Engineering, Imperial College London, London, SW7 2AZ, United Kingdom}
\author{Branko Bijeljic}
\email{b.bijeljic@imperial.ac.uk}
\affiliation[Imperial College London]
{Department of Earth Science and Engineering, Imperial College London, London, SW7 2AZ, United Kingdom}

\title[An \textsf{achemso} demo]
  {Pore-Scale Dynamics of Multiphase Reactive Transport in Water-Wet Carbonates under CO$_2$-Acidified Brine Injection: Dissolution Patterns and Reaction Rates}


\begin{document}
\pagenumbering{gobble} 
\newpage
\section*{Highlights}
\begin{itemize}
    \item  Heterogeneity of pore structure, residual oil distribution and oil re-mobilization control dissolution patterns and effective reaction rates.
    \item  Oil displacement  enhances  dissolution by channel widening, which is amplified at higher flow rates.
    \item  Reaction rates in two-phase flow are transport-limited and lower than single-phase and batch reaction rates.
\end{itemize}


\newpage
\pagenumbering{arabic}
\begin{abstract}

Depleted carbonate hydrocarbon reservoirs are promising sites for geological CO$_2$ storage, yet the presence of residual hydrocarbons introduces complex pore-scale interactions that influence the dynamics of solid dissolution. This study reveals how residual oil affects dissolution patterns and effective reaction rates during CO$_2$-acidified brine injection into Ketton limestone under reservoir conditions. We combine time-resolved X-ray microtomography (micro-CT), core-flooding experiments, and direct numerical simulations to assess the impact of pore space heterogeneity, oil distribution and injection rate. We find that the coupling between pore structure, residual oil saturation and oil displacement control flow heterogeneity, reactive surface accessibility, dissolution patterns and the reaction rates. At low injection rate, dissolution by channel widening is enhanced by oil displacement. This mechanism is especially important when dissolution is suppressed by heterogeneity in the pore space and the residual oil. At high injection rates, a more uniform dissolution occurs and can be enhanced by re-mobilisation of oil blocking brine flow. Effective reaction rates in two-phase flow are found to be lower than in the the equivalent single-phase case and up to two orders of magnitude lower than the batch rates due to persistent transport limitations. These findings offer mechanistic insights into multiphase reactive transport in carbonates and highlight the need for accurate understanding of the impact of the hydrocarbon phase on reaction to improve predictions of CO$_2$ storage efficiency.
 
\end{abstract}

\noindent \textbf{Keywords:} Reactive Transport, Muli-phase Flow, Pore Scale Imaging, CO$_2$ Storage.

\noindent \textbf{Synopsis:}  
Securely storing CO$_2$ in underground carbonate rocks is environmentally critical. This research reveals that multi-phase dynamics significantly affect effective reaction rates, a key finding for assessing the long-term stability of geological storage sites.

\newpage

\section{1. Introduction}
Carbon dioxide storage in geological formations is a key strategy for mitigating atmospheric CO$_2$ emissions~\cite{chai2025multiphase,chai2022formation, leung2014overview, bui2018carbon}. Among various geological targets, depleted carbonate oil reservoirs offer promising potential for CO$_2$ storage due to their large pore volumes, existing infrastructure, and favorable mineralogy~\cite{zhou2024review}. However, the long-term environmental security of stored CO$_2$ remains a critical concern~\cite{d2010carbon}, as interactions between injected CO$_2$, formation brine, and carbonate minerals can lead to mineral dissolution and precipitation. Specifically, the formation of carbonic acid upon CO$_2$ dissolution in brine initiates geochemical reactions with rock, altering pore structures and potentially impacting reservoir integrity~\cite{sun2023review}. These coupled fluid–rock interactions affect key properties such as porosity, permeability, and flow pathways, and must be understood to predict reactive transport behavior and ensure the stable long-term containment of CO$_2$ in carbonate formations. 

Key insights for characterizing reactive transport include identifying dissolution regimes and quantifying reaction rates. To understand dissolution regimes, a useful framework is the Péclet (Pe) and Damköhler (Da) diagram \cite{golfier2002ability, battiato2011applicability}, where Pe is the ratio of advective to diffusive rates of transport, and Da is the ratio of reaction to advection rates. In fluid–solid systems, three regimes are commonly observed. First, when both Da and Pe are high, both reactions and injection rate are rapid, leading to the fast creation of wormholes with a sharp dissolution front. This is the wormholing regime, as commonly exemplified in acidizing in oilfield operations~\cite{esteves2020pore, furui2022phase, soulaine2017mineral}. Next, when Pe is low and diffusion dominates, face (compact) dissolution occurs: a slowly advancing dissolution front progresses in the flow direction. Third, when Da is low and Pe is high, dissolution is uniform: flow reaches most reactive surfaces, reactions are slow, and material dissolves evenly throughout the volume.  However, as discussed next, whether uniform dissolution develops depends on flow heterogeneity and transport conditions.

In CO$_2$ storage, carbonic acid forms when supercritical CO$_2$ dissolves in brine. At typical reservoir conditions, its pH is about 3.1–3.5 \cite{menkeDynamicThreeDimensionalPoreScale2015}, which corresponds to a low Da and might suggest a uniform dissolution regime. 
Experiments injecting single-phase, CO$_2$-acidified brine into calcite \cite{menkeReservoirConditionImaging2016, menkeDynamicReservoirconditionMicrotomography2017} and dolomite \cite{al-khulaifiReactionRatesChemically2017a, al-khulaifiPoreScaleDissolution2019} show that low reaction rates combined with heterogeneous flow produce preferential pathways that widen over time. Channel widening then becomes the dominant dissolution mechanism. 
The occurrence and extent of channel widening depends on the initial flow field heterogeneity and the range of Pe in which both advection and diffusion are important, i.e., moderate Pe.
In contrast, more uniform dissolution emerges only at high Pe, regardless of the initial heterogeneity \cite{al-khulaifiReservoirconditionPorescaleImaging2018}.
Furthermore, the development of flow instabilities contributing to channel formation and widening has been widely reported~\cite{fredd2000advances, cooper20234d}, and shown to alter the conventional porosity–permeability relationship \cite{noirielInvestigationPorosityPermeability2004, noiriel2009changes,noirielHydraulicPropertiesMicrogeometry2005, luquotExperimentalDeterminationPorosity2009, smith2013evaporite, ottWormholeFormationCompact2015, luhmann2014experimental, yang2020dynamic}. These studies collectively highlight the critical roles transport limitations, coupled flow-dissolution dynamics and mineral heterogeneity in shaping reactive dissolution under single-phase flow conditions.

The primary knowledge gap, and the focus of this work, emerges when considering the more realistic multiphase conditions of depleted reservoirs, which invariably contain residual oil. The presence of a second fluid phase creates more complex local flow fields and may reduce the available reactive surface area~\cite{akindipePoreMatrixDissolution2022}. Previous studies on multiphase flow systems have often been decoupled, failing to capture the concurrent interactions between oil, acidic brine, and rock~\cite{singhPartialDissolutionCarbonate2018, Honarvarenergyfuels}. 
These complexities underline a fundamental difference between single-phase and multiphase flow systems: the oil phase introduces spatial and temporal heterogeneity in transport and reaction that has not yet been fully understood at the pore scale. Therefore, a complete understanding of how residual hydrocarbons influence reactive transport in carbonates is still lacking. A further challenge is the accurate estimation of effective reaction rates, which are known to be scale-dependent and significantly lower in real porous media than intrinsic rates measured in batch reactors~\cite{white2003effect, maher2010dependence, menkeDynamicThreeDimensionalPoreScale2015}.
This discrepancy is further influenced by mineral reactivity (e.g., dolomite vs. calcite), rock heterogeneity, and flow conditions including the Péclet number~\cite{peng2015kinetics, menkeDynamicThreeDimensionalPoreScale2015, al-khulaifiReactionRatesChemically2017a, al-khulaifiReservoirconditionPorescaleImaging2018}.

To address this knowledge gap, the present study investigates multiphase reactive transport in carbonate rocks under reservoir-relevant conditions using an integrated experimental and modeling approach. We design the experiments to inject CO$_2$-saturated brine for two Pe into two Ketton limestone samples containing varying levels of remaining oil. Time-resolved X-ray microtomography is used to capture dynamic changes in pore structure and fluid occupancy. Based on these 4D datasets, we perform direct numerical simulations \cite{bijeljicInsightsNonFickianSolute2013,bijeljicPredictionsNonFickianSolute2013} to quantify the evolution of porosity, permeability, velocity fields, and reactive surface accessibility. This approach allows us to systematically evaluate how initial pore structure and oil saturation interact to govern flow heterogeneity, dissolution patterns, reaction rates and dissolution dynamics.

\section{2. Materials and Methods}

\subsection{2.1 Materials}

Two cylindrical samples of Ketton limestone (\SI{12}{\milli\meter} in length, \SI{6}{\milli\meter} in diameter) were studied. The rock was composed of \SI{99.1}{\percent} calcite and characterized by well-connected pores with a bimodal size distribution \cite{patmonoaji2025differential}. 
Decane was selected as the oil phase due to its well-characterized interfacial properties and frequent use in multiphase transport studies. This choice also reduces oil-rock interactions observed with crude oil \cite{akindipePoreMatrixDissolution2022}. The synthetic brine consisted of 5~wt\% \ce{NaCl} and 1~wt\% \ce{KCl} dissolved in deionized water, doped with 30~wt\% potassium iodide (\ce{KI}) to enhance X-ray contrast during imaging \cite{lin2021drainage}.
At the experimental conditions of \SI{50}{\celsius} and \SI{8}{\mega\pascal}, the viscosities of \ce{CO2} and brine were $0.020 \pm 0.001$~mPa.s \cite{fenghour1998viscosity} and $0.60 \pm 0.05$~mPa.s\cite{abdulagatov2006viscosity}, respectively, with an interfacial tension of approximately \SI{35}{\milli\newton\per\meter} \cite{chalbaud2009interfacial}.

\subsection{2.2 Experimental Methods}
\label{subsec:2.2}

\textbf{i. Preparation of CO\textsubscript{2}-saturated brine.}
The brine was equilibrated with CO\textsubscript{2} in a high-pressure reactor at 8~MPa and 50~\textdegree C for one week.

\begin{figure}[H]
  \centering
  \includegraphics[width=1\textwidth]{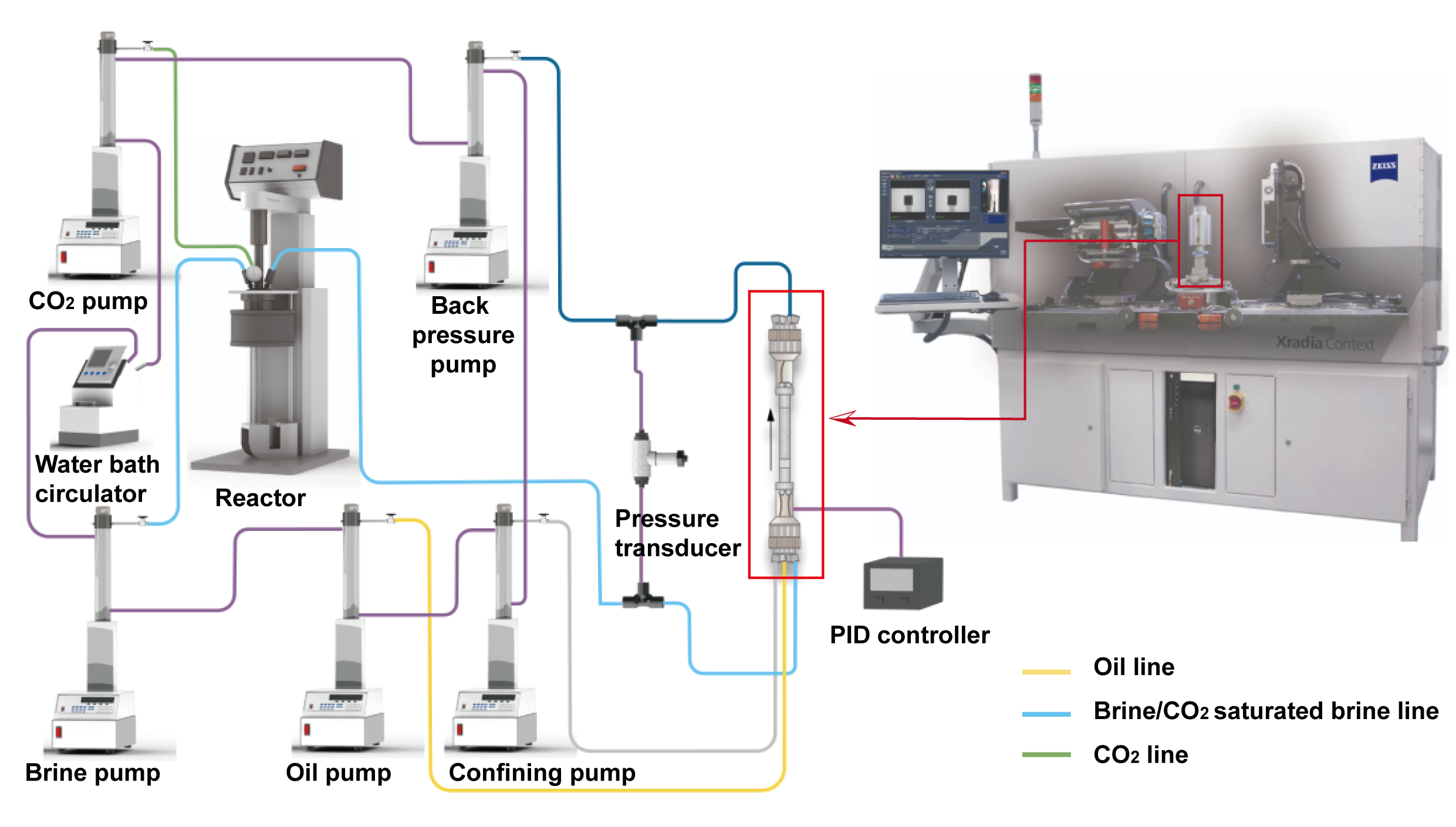}
 \caption{The experimental apparatus including flow loop and micro-CT scanner.}
  \label{fig:1}
\end{figure}

\textbf{ii. System assembly.}
The Ketton limestone sample was enclosed in a Viton sleeve and mounted into a core holder. The assembled core holder was then installed in a CT scanner and connected to the fluid delivery and receiving system, pressure transducer, and other experimental components, as shown in Figure~\ref{fig:1}.

\textbf{iii. System cleaning.}
 A confining pressure of 2~MPa was applied to ensure proper sealing and eliminate bypass flow between the sample and sleeve. The temperature was then increased to 50$^\circ$C. CO$_2$ was injected for 30~minutes to flush residual fluids and fines, followed by 24~hours of vacuuming to evacuate remaining gases.

\textbf{iv. Baseline imaging.}
A high-resolution dry scan was acquired to serve as a baseline for subsequent comparisons.

\textbf{v. Brine saturation.}
The sample was saturated with brine at a constant flow rate of 0.05~mL\,min$^{-1}$ for 100 pore volumes (PVs). The back pressure and confining pressure were then steadily increased to 8~MPa and 10~MPa, respectively, and the temperature was maintained at 50~$^\circ$C. 

\textbf{vi. Establishment of initial oil saturation (\(S_{\mathrm{oi}}\)).}
Decane was injected at 0.05~mL/min for 100 PVs to establish \(S_{\mathrm{oi}}\), after which a corresponding micro-CT scan was obtained.

\textbf{vii. Establishment of residual oil saturation (\(S_{\mathrm{or}}\)).}
Brine was re-injected at 0.05~mL/min for 20~PVs to achieve \(S_{\mathrm{or}}\), followed by another scan. The resulting \(S_{\mathrm{or}}\) values were 39.8\% for Sample~1 and 31.1\% for Sample~2.

\textbf{ix. Reactive flooding with CO\textsubscript{2}-saturated brine.}
(i) Injection from the bottom inlet at 0.05~mL/min for 315~min. 
(ii) Flow was halted for 135~min to promote diffusion-dominated transport and reaction.
(iii) Resumed injection at 0.5~mL/min for an additional 225~min.           

\subsection{2.3 Image acquisition}
To characterize the evolution of fluid occupancy and pore structure, a series of high-resolution micro-CT scans were conducted throughout the experiment. Prior to flooding, a dry scan of the sample was acquired to capture the initial pore geometry. This was followed by a brine-saturated scan after full saturation and a scan after decane injection to determine the initial oil distribution. All these pre-reaction scans used 2601 projections with an energy of \SI{100}{\kilo\electronvolt} and a power of \SI{9}{\watt}. The scans were conducted at a voxel resolution of \SI{4}{\micro\meter} and required approximately \SI{12}{\hour} each to ensure optimal image quality and segmentation accuracy.

After conducting a scan following brine re-injection to establish the residual oil configuration, time-resolved imaging commenced at the onset of CO\textsubscript{2}-saturated brine injection, designated as \(t = 0\). A total of 12 images were acquired during the course of the reactive transport experiment. Each scan was performed using 901 projections at an energy of \SI{100}{\kilo\electronvolt} and a power of \SI{9}{\watt}, with a voxel size of \SI{6}{\micro\meter} and a scan time of approximately \SI{45}{\minute}. Thus, the first scan was performed from \(t = 0\) to \(t =\) \SI{45}{\minute}: this is called the \SI{45}{\minute} scan. The second image is called \SI{90}{\minute} and so on. This imaging protocol was selected to balance temporal resolution with image quality, allowing for accurate tracking of dissolution fronts and structural evolution without significant motion artifacts or radiation damage.

\subsection{2.4 Image Processing and Analysis}

All raw micro-CT datasets were processed using a standardized procedures comprising denoising, normalization, registration, and resampling, thereby ensuring uniform voxel resolution and spatial alignment across acquisition time points. Phase segmentation employed a hybrid workflow that combined differential imaging, interactive thresholding, and watershed-based segmentation~\cite{Lin2016,chai2025multiphase}, enabling robust delineation of three primary phases: rock matrix, brine, and oil.

\textbf{Pre-reaction datasets.}
(i) The rock matrix was extracted from the dry reference scan.
(ii) Oil was segmented directly from the images without differential imaging.
(iii) Brine was isolated by subtracting the dry scan from the brine- or oil-saturated scans, thereby identifying the pores occupied by the aqueous phase.

\textbf{Post-reaction datasets.}
(i) Mineral dissolution was quantified by subtracting the segmented pre-reaction rock volume from the post-reaction image, yielding the dissolved fraction.
(ii) The reacted rock matrix was obtained by subtracting the dissolution volume from the original rock phase.
(iii) Brine and oil were segmented using the same protocol as for the pre-reaction datasets, ensuring methodological consistency across all time steps.

\subsection{2.5 Pore-Scale Simulation}
Following image segmentation, the lower region of the segmented micro-CT volume was selected for pore-scale flow simulation in the brine phase. The modeling approach follows the method developed by \citet{bijeljicInsightsNonFickianSolute2013} and \citet{RAEINI20125653}, implemented within the \texttt{OpenFOAM} framework. The solver employs the finite volume method to simultaneously solve the continuity and steady-state incompressible Navier–Stokes equations:
\begin{align}
\nabla \cdot \mathbf{u} &= 0 \\
\rho \left( \frac{\partial \mathbf{u}}{\partial t} + \mathbf{u} \cdot \nabla \mathbf{u} \right) &= -\nabla p + \mu \nabla^2 \mathbf{u}
\end{align}
where $p$ is the pressure ($Pa$), and $u$ is the velocity ($m/s$), both obtained for each voxel of the image; $\mu$ is the fluid (brine) viscosity ($Pa\cdot s$); $\rho$ is the fluid density ($kg/m^3$).
The flow rate (\( \text{m}^3/\text{s} \)) is calculated as \( Q = \int u_x \, dA_x \), where \( A_x \) is the cross-sectional area of the image (\( \text{m}^2 \)) and \( u_x \) is the velocity in the direction of overall flow (\( \text{m}/\text{s} \)). The Darcy velocity is then calculated as \( u_D = \frac{Q}{L_y L_z} \), where \( L_y \) and \( L_z \) are the lengths of the image (\( \text{m} \)).

A fixed pressure drop was applied along the flow direction, while all solid boundaries were treated as no-slip walls. To analyze the spatial distribution of flow, the velocity field was normalized by the average pore velocity:
\begin{equation}
U_{\text{av}} = \frac{u_D}{\phi}
 \label{Eq:Uav}
 \end{equation}
where \( \phi \) is the porosity (dimensionless), and the ratio \( u / U_{\text{av}} \) was visualized to assess local deviations from the mean flow behavior.

\subsection{2.6 Dimensionless numbers and reaction rates}
In this study, the dimensionless number, Péclet (Pe) and Damköhler (Da), are used to characterize and quantify the reactive transport.
The Péclet number quantifies the relative efficiency of solute mass transfer through advection compared to diffusion \cite{peclet1827traite}:
\begin{align}
   \text{Pe} = \frac{\text{advective transport rate}}{\text{diffusive transport rate}} = \frac{u_{\text{avg}} L_c}{D_m}
   \label{Pe}
\end{align}
where $D_m$ is molecular diffusion coefficient in brine ($m^2/s$), and $L_c$ is characteristic length ($m$), calculated by \cite{mostaghimi2012simulation}:
\begin{align}
    L_c = \frac{\pi}{S}
\end{align}
where the specific surface area S (m$^{-1}$), is the image surface area per unit volume at the beginning of the time period, calculated by $V_B/As$, where $V_B$ is bulk volume, and $As$ is the surface area from image analysis.

The Damköhler number quantifies the ratio between the timescales of a chemical reaction and the mass transfer \cite{lasaga1984chemical}:
\begin{align}
    \text{Da} = \frac{\text{reaction rate}}{\text{advective transport rate}} = \frac{L_c}{u_{\text{avg}}} k
    \label{12}
\end{align}
where $k$ is the chemical reaction rate constant (s$^{-1}$), calculated by: $ k = \frac{\pi r}{nL}$.
r is the mineral reaction rate (\(\text{mol}\cdot\text{m}^{-2} \cdot \text{s}^{-1}\)), L is the sample length ($m$), and $n$ is calculated by \cite{menkeDynamicThreeDimensionalPoreScale2015}:
\begin{align}
    n = \frac{\rho_{\text{mineral}} f_{\text{mineral}}}{M_{\text{mineral}}}
\end{align}
where $\rho_{\text{}}$ is the mineral density, $M_{\text{mineral}}$ is their molecular mass. Therefore, equation \ref{12} can be rewritten as follows\cite{al-khulaifiReservoirconditionPorescaleImaging2018}:
\begin{align}
    \text{Da} = \frac{\pi r_{\text{mineral}}}{u_{\text{avg}} n}
    \label{Da}
\end{align}
where $r_{\text{}}$ is the non-transport limited reaction rate.

The effective reaction rate ($r_{\text{eff}}$) of mineral is determined as \cite{al-khulaifiReactionRatesChemically2017a}:
\begin{align}
    r_\text{eff} = \frac{\rho_\text{mineral} (1 - \phi_\text{grain}) \Delta \phi_\text{CT}}{M_\text{mineral} S \Delta t}
\end{align}
where $\Delta t$ is the time between scans (s), $\Delta \phi_\text{CT}$ is the corresponding change in porosity and $S$ is the image surface area per unit volume (\(\text{m}^{-1}\)). The term $1 - \phi_\text{grain} = 0.88$ accounts for the fact that the grains themselves have micro-porosity \cite{menkeDynamicThreeDimensionalPoreScale2015}.

\section{3. Results and Discussion}
\subsection{3.1 Pore-Scale Heterogeneity}
\label{subsec:3.1}
The interplay between initial pore-scale structural heterogeneity and the subsequent distribution of residual oil creates a coupled dual heterogeneity, which governs the flow field and transport dynamics. Our findings reveal that although both carbonate samples originate from the same formation, their distinct pore structure leads to different distributions of residual oil and, consequently, contrasting dissolution patterns and reaction rates.

\begin{figure}[H]
  \centering
  \includegraphics[width=1\textwidth]{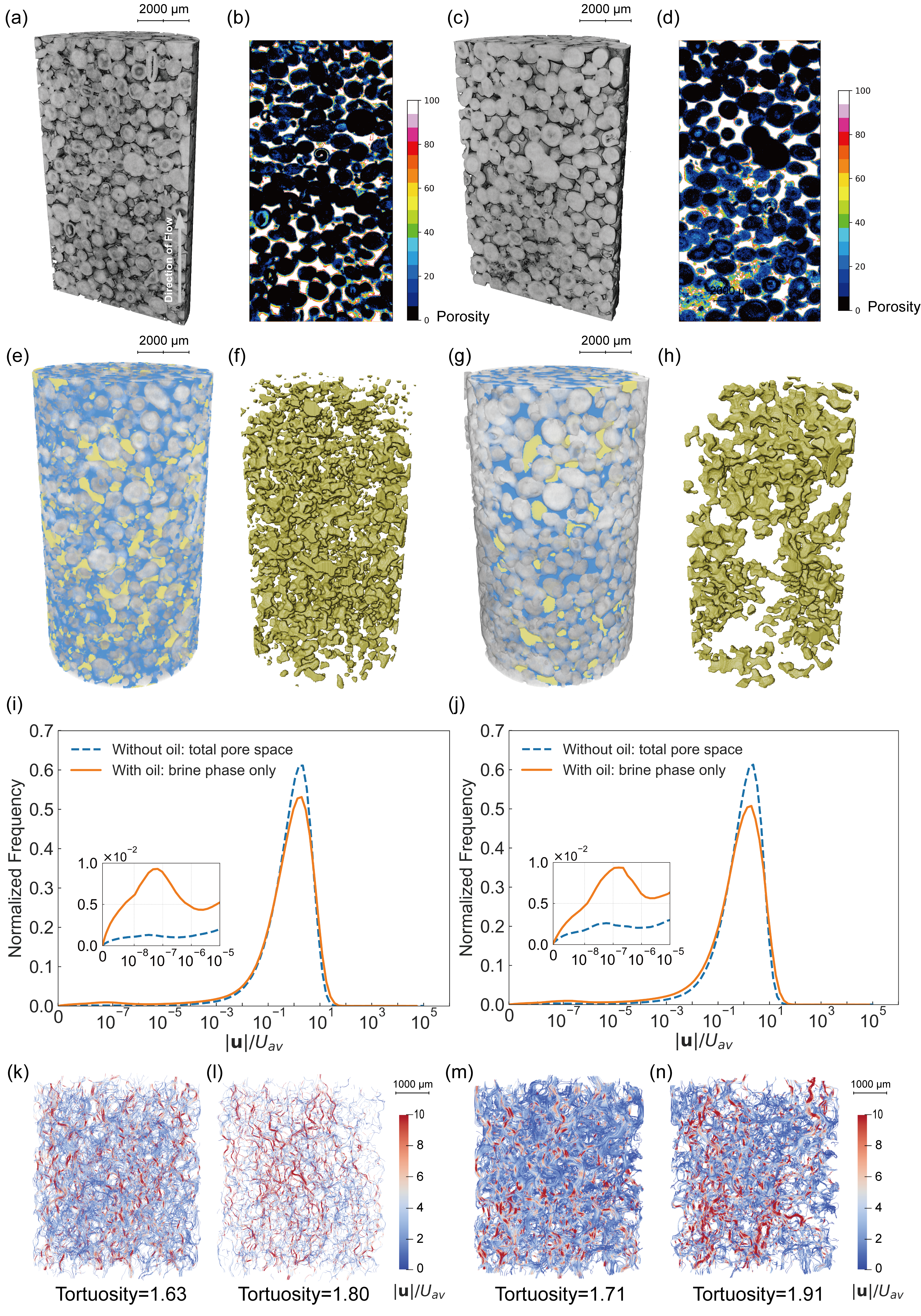}
 \caption{ 3D dry-image renderings for Sample 1 (a) and Sample 2 (c) show solids in gray. Porosity maps (b, d) color unresolved porosity: white = resolved macro-pores (porosity = 1), black = solid (0), intermediate shades = unresolved pores. Fluid distributions (e–h) show oil in yellow, brine in blue and solids in gray. Initial PDFs of $\log$-velocity compare with/without oil for Sample 1 (i) and Sample 2 (j); voxel velocities $u$ are sorted into 128 bins of $\log(|u|/U_{\text{av}})$, where $|u|/U_{\text{av}}$ is the ratio of $|u|$ to the average velocity defined by Eq.~\ref{Eq:Uav}. Streamlines show velocity fields without and with oil for Samples 1 (k–l) and 2 (m–n), with corresponding tortuosities. }
  \label{fig:2}
\end{figure}

Sample 1 is characterized by a well-connected network of intergranular macro-pores, resulting in a relatively homogeneous structure (Figures~\ref{fig:2}a, b). This architecture facilitates more uniform brine flooding, leading to a piston-like displacement and a relatively dispersed distribution of residual oil throughout the accessible pore space (Figures~\ref{fig:2}e, f). Based on SEM and mercury intrusion data \cite{patmonoaji2025differential}, Sample 2's heterogeneity stems from a dual-porosity system consisting of intergranular macro-pores and poorly-connected intermediate sized pores formed by fine interstitial particles (Figures~\ref{fig:2}c, d), in contrast to Sample 1's more uniform macro-pore network. The porosity map (Figure~\ref{fig:2}b and d) further reveals that Sample 2 exhibits significant intergranular packing of interstitial particles, particularly near the inlet, which appears as intermediate gray tones. During brine flooding, this structural complexity promotes capillary fingering and incomplete displacement, causing residual oil to become concentrated in localized, isolated clusters (Figures~\ref{fig:2}f, h). The resulting oil clusters in Sample 2 act as localized flow barriers, further amplifying the intrinsic structural heterogeneity.

To quantify the impact of this dual heterogeneity on transport, pore-scale simulations were performed, the results of which underscore the compounding effect of structure and saturation. The flow field was computed in the pore space containing brine (with oil) and where we assumed that no oil was present, so the flow field was computed in the entire pore space (without oil). In the absence of oil, both samples show relatively narrow, unimodal velocity probability density functions (PDFs) indicative of more uniform flow, although Sample 2 already displays a slightly higher intrinsic tortuosity due to its more complex intergranular pore structure (eg., 1.63 vs 1.71) (Figure~\ref{fig:2}k, m). As shown in Figure~\ref{fig:2}i and j, the probability density functions (PDFs) of velocity in the without oil cases display narrow, unimodal distributions centered around the average pore velocity, indicating relatively uniform flow. In contrast, in the with oil cases, the PDFs are broader with reduced peak velocities and a marked increase in the fraction of low-velocity voxels, especially in Sample 2. This is visually confirmed by the 3D velocity renderings (Figure~\ref{fig:2}k--n), which reveal that residual oil forces the flow into highly disconnected and tortuous pathways in Sample 2, while flow in Sample 1 remains better connected. Quantitatively, this translates to a higher tortuosity of 1.91 in the with oil case for Sample 2 compared to tortuosity of 1.80 in Sample 1. These results provide quantitative evidence that dual heterogeneity—stemming from both intrinsic pore structure and residual oil distribution—amplifies transport heterogeneity by disrupting flow continuity and enhancing the tortuosity of the local flow paths.

In summary, our results demonstrate that flow field heterogeneity is not merely a function of the rock's static pore structure but is amplified by the distribution of the residual phase. Sample 2, with its high intrinsic structural heterogeneity, fosters a trapping mechanism that creates a higher layer of heterogeneity from the residual oil. This dual system synergistically enhances flow tortuosity and reduces overall permeability, creating diffusion-limited zones that are likely to significantly influence reactive transport and mineral dissolution patterns, as explored in Section 3.2\ref{subsec:3.2}. Moreover, we will see in the next section how widening of the pore space as a consequence of displaced oil (coupling of rock dissolution and oil displacement) has a profound impact on dissolution patterns.

\subsection{3.2 CO$_2$-Acidified Brine Injection and Dissolution}
\label{subsec:3.2}
Reactive transport experiments with CO$_2$-saturated brine revealed two distinct dissolution patterns governed by the interplay between initial pore-scale heterogeneity and multiphase displacement dynamics. Sample 1 developed dominant, displacement-driven channels, whereas Sample 2 showed dissolution suppressed by its fine-scale intergranular packing and oil blockage.

\begin{figure}[H]
  \centering
  \includegraphics[width=1\textwidth]{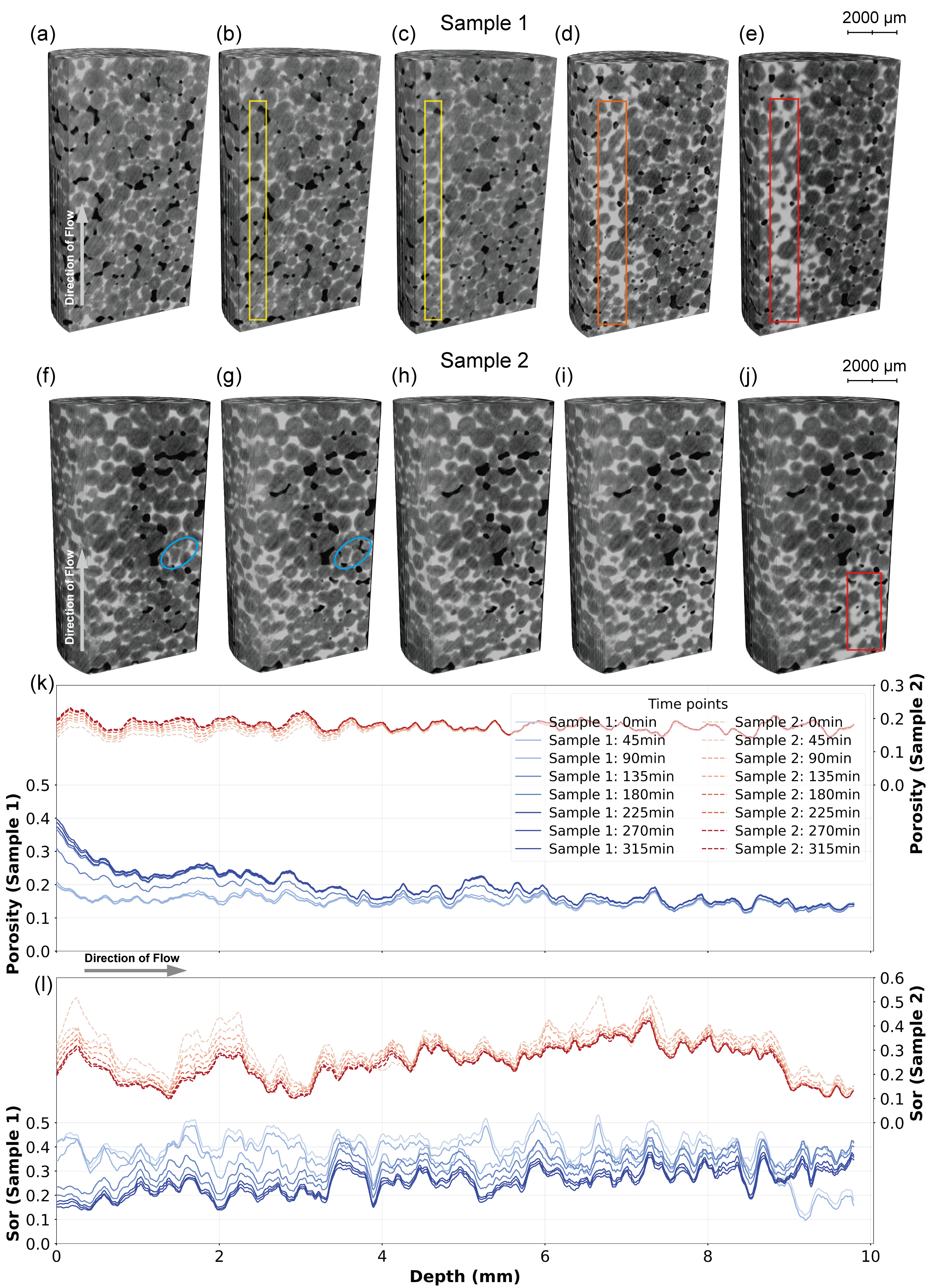}
 \caption{ Grayscale images depicting the temporal evolution of dissolution in Sample 1 (a–e) and Sample 2 (f–j) at 0 minutes, 45 minutes, 90 minutes, 135 minutes and 315 minutes, under a flow rate of 0.05 ml/min. In these images, gray represents solids, black represents oil, and white represents brine. Temporal evolution of porosity and oil saturation averaged across slices perpendicular to the flow direction. (k): Porosity profiles for Sample 1 and Sample 2. (l): oil saturation profiles for Sample 1 and Sample 2. As rock dissolution advances, porosity increases while oil saturation decreases. The \SI{0}{\minute} mark corresponds to the scan before reaction. The \SI{45}{\minute} mark corresponds to the first scan performed from \(t = 0\) to \(t =\) \SI{45}{\minute}, and \SI{90}{\minute} corresponds to the second scan performed from \(t = 45\) to \(t =\) \SI{90}{\minute}, and so on for subsequent scans.}
  \label{fig:3}
\end{figure}

The channel formation and widening in Sample 1 is a direct consequence of multiphase flow dynamics; the presence of oil and its displacement creates a more heterogeneous flow field and encourages channel formation compared to single-phase flow \cite{menkeReservoirConditionImaging2016, menkeDynamicReservoirconditionMicrotomography2017, al-khulaifiReactionRatesChemically2017a, al-khulaifiPoreScaleDissolution2019}. The process initiates in the larger pores where the injection of brine initially causes little dissolution because the presence of oil hinders the flow of brine. As the pore space is altered by the initial dissolution, preferential pathways are progressively widened and then further expanded by a displacement-induced oil re-mobilization mechanism. This pore-scale process is directly visualized in its early stages (Figure~\ref{fig:3}b-c, yellow box), where the brine dissolved the confining carbonate pore throats and allowing for trapped oil to be displaced. This action triggers a positive feedback loop: oil displacement improves acid access, which in turn accelerates dissolution and further enhances channel breakthrough and widening. This self-reinforcing process culminates in the channel breaking through the entire sample by 135 minutes (Figure~\ref{fig:3}d, orange box). This event is corroborated by the sharp, concurrent increase in porosity and decrease in $S_{\mathrm{or}}$ at 135 minutes, as shown in the profiles (Figure~\ref{fig:3}k-j). Ultimately, this leads to the formation of a single, highly conductive dominant channel by 315 minutes (Figure~\ref{fig:3}e, red box). The importance of oil re-mobiblization can be seen in residual oil saturation profiles. In Figure~\ref{fig:3}j, at the outlet of Sample 1, the oil saturation initially increased but then decreased. This is due to an oil bank flowing through the sample but getting stuck at the outlet between 45 and 90 minutes, which subsequently gets released.

Conversely, the heterogeneity of the initial pore space and residual oil distribution in Sample 2 effectively suppressed this feedback loop. 
Here, the presence of fine intergranular particles in the region near the inlet (Figure~\ref{fig:2}c-d) leads to high capillary entry pressures and reduced flow connectivity. Therefore, the channel formation caused by displacement is confined to a limited region. However, in this region, the blockage of flow pathways by oil hinders further channel development and breakthrough. This is reflected in the porosity and $S_{\mathrm{or}}$ profiles, which show only modest and localized changes at the inlet, confirming a spatially confined reaction. Comparing brine to oil occupying the blue circles A (0 minutes) and B (45 minutes) in Figure~\ref{fig:3}d showing the gray scale images of Sample 2, we can also clearly see the small oil ganglion that blocks channel formation from 45 minutes (Figure~\ref{fig:3}d, blue circle). This results in the channel formation being confined to this inlet region (Figures\ref{fig:3}j, red box) preventing feedback-driven channel development.

In conclusion, this analysis elucidates how initial pore-scale heterogeneity dictates dissolution patterns within a multiphase flow context. The results distinguish between two phenomena: 1) Displacement-driven channel growth and widening, with dissolution-induced further displacement to create a positive feedback loop; and 2) heterogeneity and oil blockage-suppressed dissolution, where fine-scale features in the low-permeability region limit channel formation and oil blockage prevents channel breakthrough. This demonstrates that predicting reactive transport in subsurface reservoirs requires models that account not only for static heterogeneity but also for the dynamic, pore-scale coupling between fluid displacement and mineral reaction.

\subsection {3.3 Flow Fields}
\label{subsec:3.3}
\begin{figure}[H]
  \centering
  \includegraphics[width=1.0\textwidth]{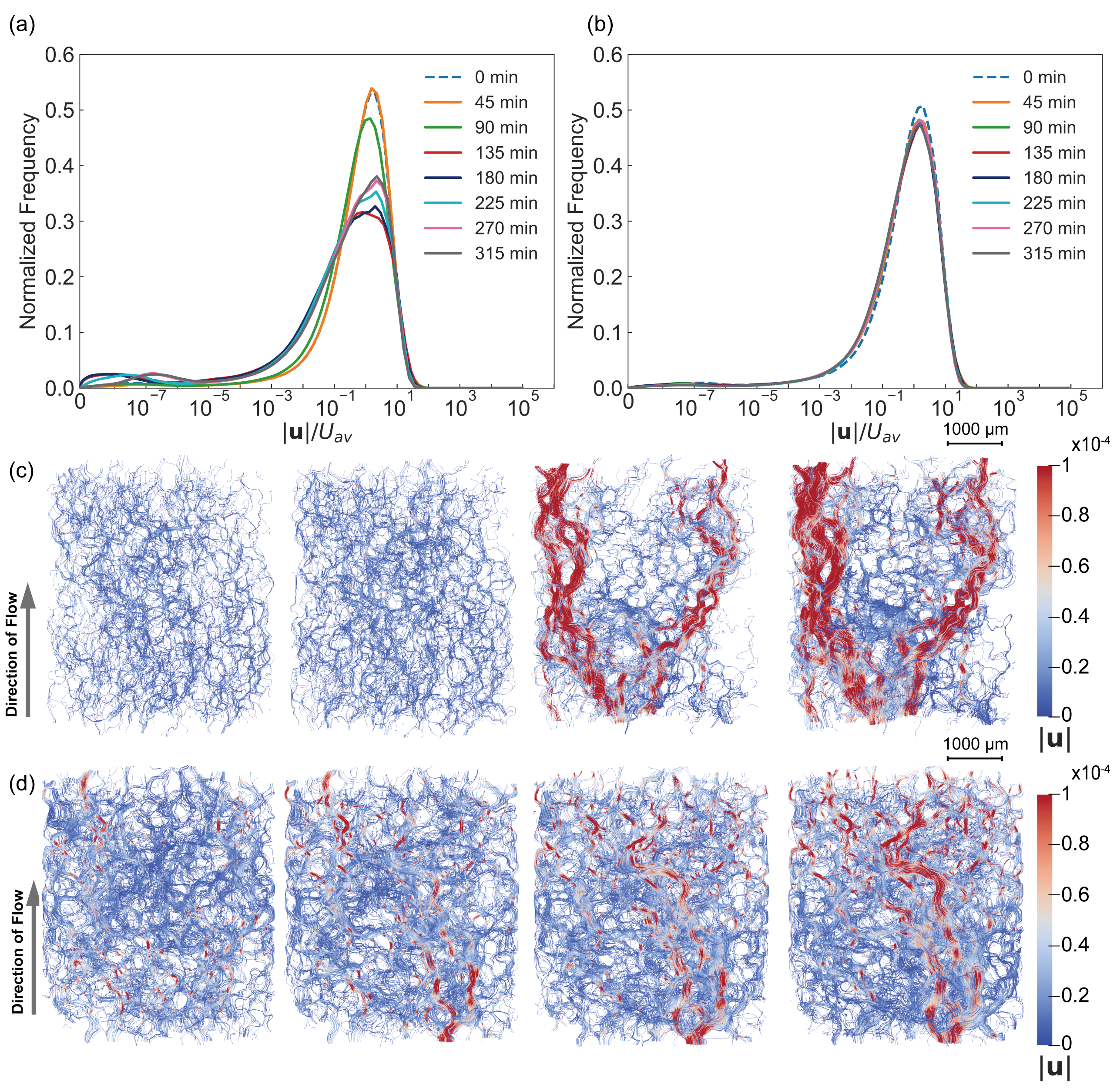}
  \caption{
 Probability density functions (PDFs) of the velocity for Sample 1 (a) and Sample 2 (b), with oil over time (from 0 minutes to 315 minutes). Voxel velocities, $u$, were collected from the simulations and uniformly sampled into 128 bins of $\log(|u|/U_{\text{av}})$, where $|u|/U_{\text{av}}$ represents the ratio of the magnitude of $u$ to the average velocity, $U_{\text{av}}$, Eq.~\ref{Eq:Uav}. Velocity streamlines for Sample 1 (c) and Sample 2 (d) at 0 minutes, 45 minutes, 135 minutes and 315 minutes. The \SI{0}{\minute} mark corresponds to the residual oil scan. The \SI{0}{\minute} mark corresponds to the scan before reaction. The \SI{45}{\minute} mark corresponds to the first scan performed from \(t = 0\) to \(t =\) \SI{45}{\minute}, and \SI{90}{\minute} corresponds to the second scan performed from \(t = 45\) to \(t =\) \SI{90}{\minute}, and so on for subsequent scans.}
  \label{fig:4}
\end{figure}

Flow modeling provides a dynamic visualization of how the two distinct dissolution patterns, established in the previous section, manifest in the fluid flow field and govern the evolution of transport properties. Analysis of the time-resolved velocity distributions immediately quantifies the contrasting flow dynamics. In Sample 1, the velocity probability density function (PDF) progressively broadens and shifts, indicating a significant increase in flow heterogeneity as the system evolves, with more stagnant velocities and fewer velocities similar to the mean velocity (Figure~\ref{fig:4}a). This is the statistical signature of channel formation and widening. The 3D velocity streamlines (Figure~\ref{fig:4}c) provide a visual confirmation, showing the formation of two high-velocity preferential flow paths that monopolize the flow, leaving large surrounding regions stagnant. In contrast, Sample 2 maintains a stable, narrow velocity PDF over time (Figure~\ref{fig:4}b), which, combined with its more spatially distributed flow field (Figure~\ref{fig:4}d), confirms a distinct time-dependent dissolution pattern initially characterized by localized dissolution near the inlet, and development of localized channel that did not fully break through.

\begin{figure}[H]
  \centering
  \includegraphics[width=0.7\textwidth]{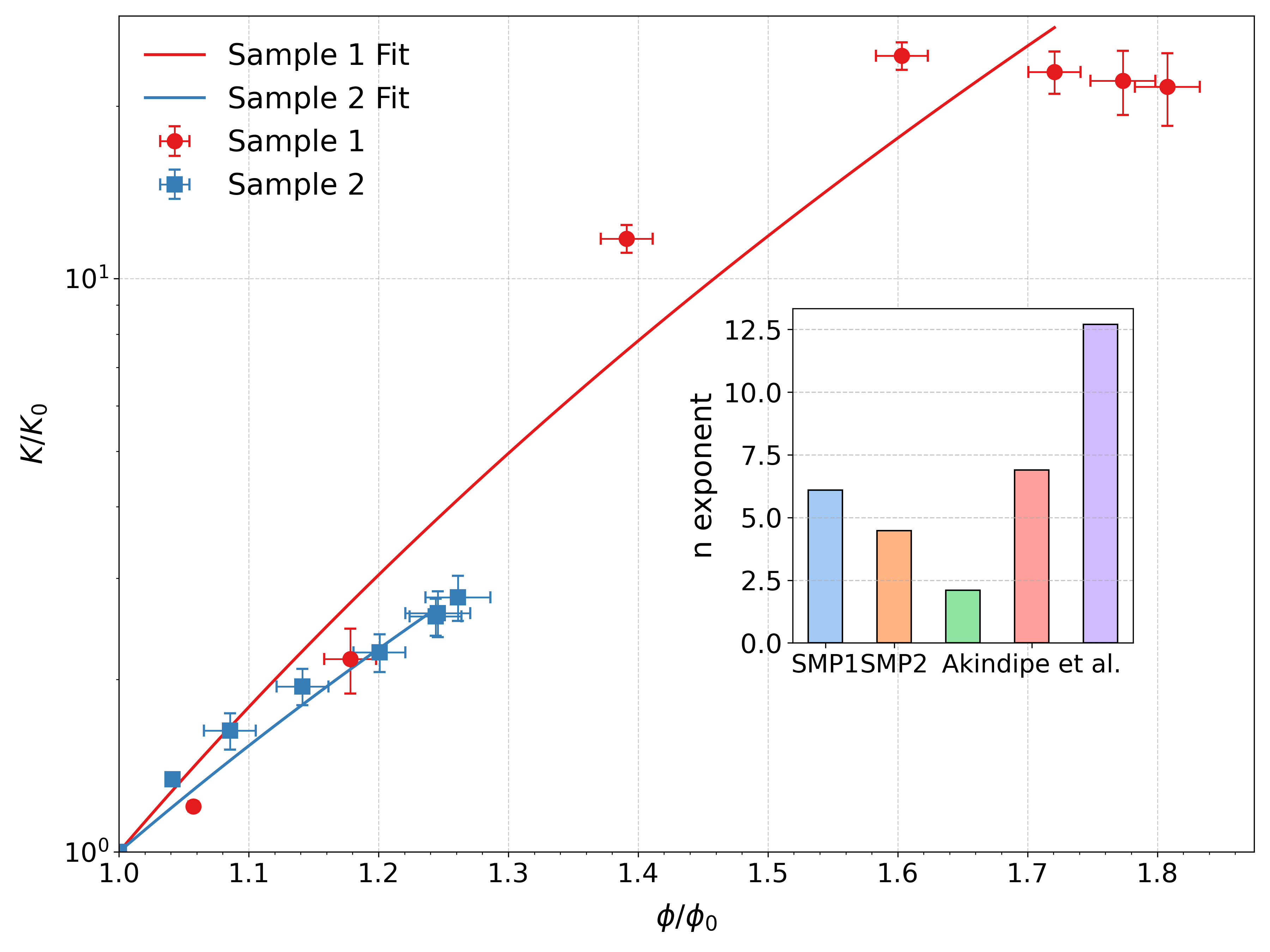}
  \caption{
The permeability-porosity relationships exhibit clear distinctions between Sample 1, characterized by channel widening, and Sample 2, which displays heterogeneity and oil blockage supressed dissolution. The best-fit exponents ($n$) in Eq.~\ref{Eq:Kphi} are  $6.1$ for Sample 1, and it is $4.5$ for Sample 2. For comparison, the inset shows the measured exponents in the work of \citet{akindipePoreMatrixDissolution2022}}
  \label{fig:5}
\end{figure}

This profound difference in flow field evolution directly dictates the macroscopic relationship between permeability ($K$) and porosity ($\phi$). The relationship between permeability and porosity for both samples can be described by a power-law relationship of the form:

\begin{equation}
\frac{K}{K_0} = \left( \frac{\phi}{\phi_0} \right)^n
\label{Eq:Kphi}
\end{equation}

\noindent where \( K \) and \( \phi \) are the permeability and porosity at a given time, \( K_0 \) and \( \phi_0 \) are their corresponding initial values, and \( n \) is the power-law exponent characterizing the sensitivity of permeability to changes in porosity.  This relationship is shown in Figures~\ref{fig:5}.

For our experiments, Sample 1 yielded n = 6.1, while Sample 2 yielded n = 4.5. The high exponent in Sample 1 confirms that the observed channel widening is a highly efficient mechanism for permeability enhancement, where dissolution creates a disproportionately large increase in flow capacity. The lower exponent in Sample 2 reflects a less efficient, spatially confined dissolution process due to a low permeability region. In previous work, \citet{akindipePoreMatrixDissolution2022}, summarized in Figure~\ref{fig:5}, reported a range of exponents for different multiphase dissolution patterns: a highly efficient, sample-spanning channel yielded n = 12.7, whereas a less stable, tapering channel gave n = 2.1. Our exponent of 6.1 for Sample 1 is comparable to the n = 6.9 reported for a similar dissolution regime. The n value of 4.5 for Sample 2 places it between the highly focused channeling of Sample 1 and more uniform dissolution reported by others, consistent with our visual evidence of a spatially confined reaction that lacks a single, dominant conduit.
Overall, Sample 1 exhibits a large, nearly two-order-of-magnitude increase in permeability. Critically, this increase is not smooth but occurs in distinct steps. These steps are the macroscopic signature of the channel breakthrough events visualized in the flow fields (Figure~\ref{fig:4}c). Once a dominant channel connects the inlet and outlet, permeability increases significantly. Conversely, Sample 2 shows a modest and continuous permeability increase, reflecting the gradual, uniform enlargement of many small pores rather than the creation of a single dominant channel.

In summary, the flow simulations provide a crucial link between the pore-scale dissolution patterns and their macroscopic consequences. They demonstrate that channel widening creates preferential flow paths, leading to a large increase in permeability. The inlet heterogeneity and oil blockage-suppressed dissolution, so the pore space alteration was not focused, resulting in a much more modest and predictable enhancement of transport properties. This highlights that for reactive transport, knowing the change in average porosity alone is insufficient to predict the change in permeability; the spatial pattern of that change is paramount.

\subsection {3.4 Effective Reaction Rates}
\label{subsec:3.4}
\begin{figure}[H]
  \centering
  \includegraphics[width=1\textwidth]{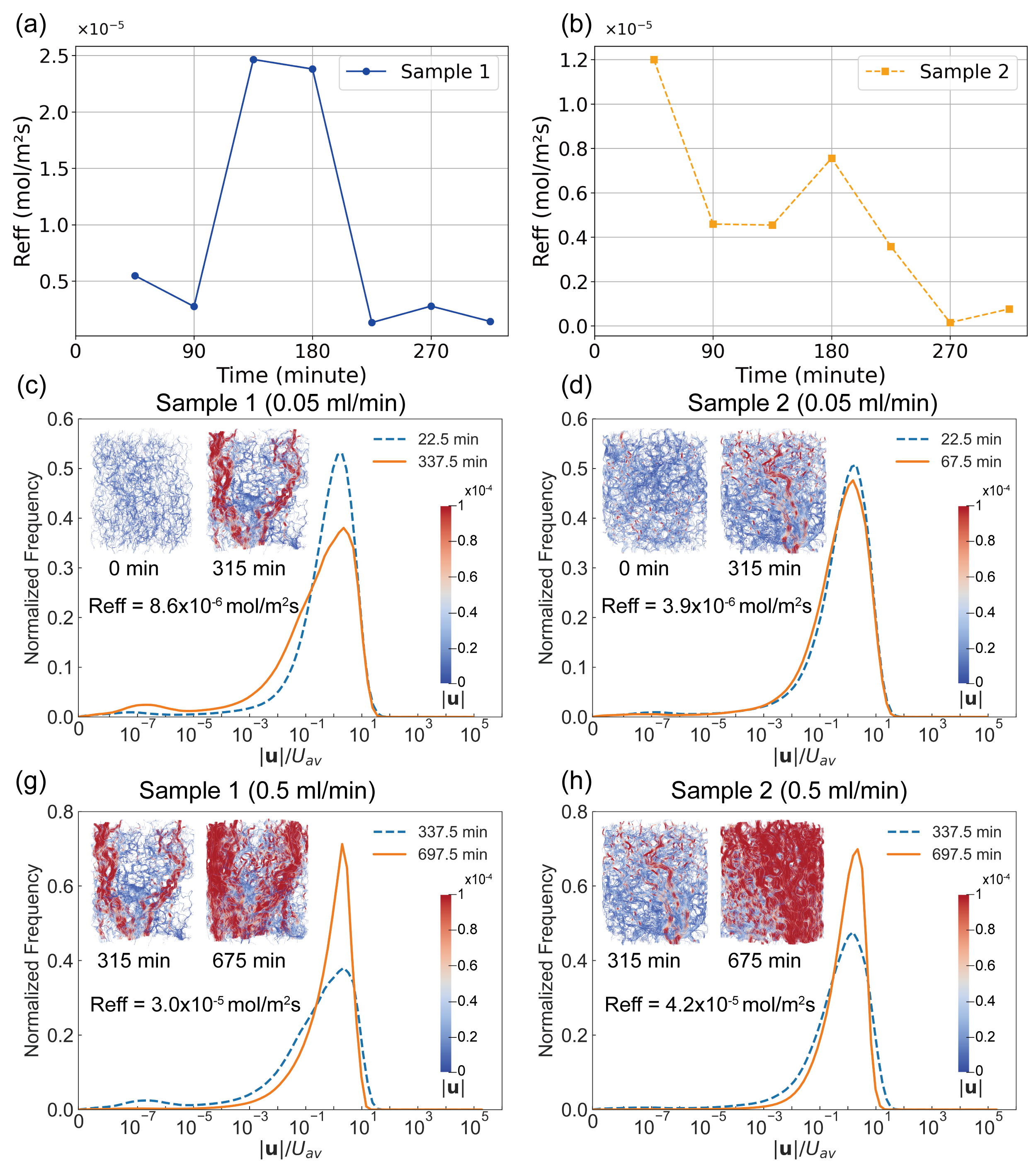}
  \caption{(a)-(b): Effective reaction rate as a function of experimental time for Samples 1 and 2 (45–315 minutes).(c)-(f): Probability density functions (PDFs) of velocity for Sample 1 and Sample 2 at 0 minutes, 315 minutes, and 675 minutes and corresponding 3D renderings of voxel velocity normalized by average velocity. The effective reaction rate at \SI{45}{\minute} corresponds to the first scan performed from \(t = 0\) to \(t =\) \SI{45}{\minute}, which represent the average state at 22.5 min (\(\Delta t=22.5\) min), and \SI{90}{\minute} corresponds to the second scan performed from \(t = 45\) to \(t =\) \SI{90}{\minute}, which represents the average state at 67.5 min (\(\Delta t=45\) min), and so on for subsequent time step} 
  \label{fig:6}
\end{figure}

The influence of dissolution patterns on reaction kinetics was investigated by comparing the two samples under two flow rates (Figure~\ref{fig:6}). At low flow rates, the samples exhibited distinct behaviors due to their structural and oil distribution differences. Sample 1 exhibited channel formation and widening as the injected acid displaced oil, resulting in a sharp initial increase in the surface-normalized reaction rate, followed by a decline (Figure~\ref{fig:6}a). This trend suggests that while channel formation initially enhances acid transport to new surfaces, their stabilization ultimately leads to inefficient mass transfer as a significant portion of the acid bypasses the bulk of the reactive matrix. In contrast, Sample 2, characterized by presence of interstitial particles in the region near the inlet, showed a lower and more stable reaction rate (Figure~\ref{fig:6}b). Its low connectivity caused rapid acid consumption only near the inlet (Figures~\ref{fig:3}j, red box), initially confining the reaction to a dispersed pattern of isolated zones and preventing the formation of a continuous, effective channel (Figures~\ref{fig:3}j).

\begin{table}[H]
\centering
\caption{Comparison of reaction rates in our multiphase flow experiments with single-phase flow experiments results in the literature and the batch reaction rate.}
\label{table1}
\begin{tblr}{
  cells = {c},
  cell{1}{2} = {c=2}{},
  cell{8}{2} = {c=2}{},
  cell{9}{2} = {c=2}{},
  hline{1-2,10} = {-}{},
}
Condition                             & Reaction Rate (mol/m\textsuperscript{2}s) &                             \\
Ketton limestone multiphase~ ~ ~ & Sample 1                                  & Sample 2                    \\
Low flow rate~                         & 8.6×10\textsuperscript{-6}                & 3.9×10\textsuperscript{-6}  \\
High flow rate                         & 3.0×10\textsuperscript{-5}                & 4.2×10\textsuperscript{-5}  \\
Silurian dolomite single-phase  \cite{al-khulaifiReservoirconditionPorescaleImaging2018}    & Heterogeneity A                           & Heterogeneity B             \\
Low flow rate                          & 3.15×10\textsuperscript{-6}               & 1.35×10\textsuperscript{-6} \\
High flow rate                         & 5.47×10\textsuperscript{-6}               & 10.6×10\textsuperscript{-6} \\
Ketton limestone single-phase \cite{menkeDynamicThreeDimensionalPoreScale2015} ~   & 5.0×10\textsuperscript{-5}                &                             \\
Batch reaction rate \cite{peng2015kinetics}                   & 6.9×10\textsuperscript{-4}                &                             
\end{tblr}
\end{table}

The difference in dissolution patterns directly translated to differing overall reaction rates. As summarized in Table~\ref{table1}, the average reaction rate in Sample 1 ($8.6 \times 10^{-6}$ mol/m\textsuperscript{2}s) was 2.2 times higher than in Sample 2 ($3.9 \times 10^{-6}$ mol/m\textsuperscript{2}s), which experienced heterogeneity and oil blockage suppressed dissolution. Our finding that channel widening leading to the new channel breakthrough in Sample 1 leads to a higher average reaction rate is conceptually consistent with the single-phase flow experiments of \citet{al-khulaifiReservoirconditionPorescaleImaging2018}. In this single-phase study, the dominant channel(s)  formation was linked to the initial physical heterogeneity and the lower Pe studied, where a more homogeneous rock structure led to a higher effective reaction rate (Table~\ref{table1}). However, the channel widening in our study is also a direct consequence of multiphase displacement dynamics. As established in Section 3.2, the dominant channel in Sample 1 is not only a feature of the pore space structure but is dynamically carved out by a positive feedback loop involving dissolution-induced oil displacement. While the channels in single-phase flow are a feature of the rock's initial physical heterogeneity, in multiphase flow they are also a feature of the residual oil distribution and the fluid-fluid displacement process. 
Nevertheless, both rates are substantially lower than those from single-phase flow\cite{menkeDynamicThreeDimensionalPoreScale2015} ($5.0 \times 10^{-5}$) or batch experiments \cite{peng2015kinetics} ($6.9 \times 10^{-4}$), highlighting the critical role of multiphase flow and pore-scale transport limitations. 

A quantitative analysis of the transport regime during the low flow rate injection further reveals its inherent limitations. As dissolution progressed, both samples exhibited a decreasing Péclet (Pe) number and an increasing Damköhler (Da) number (Appendix, Figure~\ref{fig:a1}). The decrease in Pe, caused by the relative slowing of fluid flow as pore channels widen, indicates a shift from a more advection-dominated to a more diffusion-influenced transport. Concurrently, the rise in Da—whose non-monotonic trend reflects the complex evolution of local porosity and surface area—is attributed to the growth of available reactive surfaces. The crucial finding, however, is that the combined PeDa ratio increased only slightly for both samples. This confirms that under these low-flow conditions, the rate-limiting step for the overall dissolution process was consistently pore-scale diffusive mass transport.

Increasing the injection rate ten-fold further revealed the underlying rate-limiting mechanisms. Sample 1 exhibited a sub-linear, three-fold increase in effective reaction rate (to $3.0 \times 10^{-5}$ mol/m\textsuperscript{2}s), indicating that the reaction was constrained by the established channel network; higher velocity primarily accelerated flow through existing high-permeability paths rather than engaging new reactive surfaces. Conversely, Sample 2 demonstrated a near-linear increase (to $4.2 \times 10^{-5}$ mol/m\textsuperscript{2}s), surpassing Sample 1's rate (Table~\ref{table1}). This suggests that the higher flow rate successfully overcame the initial diffusion limitations, forcing acidic brine into previously inaccessible pore spaces (including those blocked by oil) and promoting a more uniform, globally effective dissolution. This behavior is consistent with the findings by \citet{al-khulaifiReservoirconditionPorescaleImaging2018} in dolomite rock and with studies showing that increased flow can enhance permeability and redistribute flow, overcoming local reactant consumption limitations\cite{al-khulaifiReservoirconditionPorescaleImaging2018}.

\begin{figure}[H]
  \centering
  \includegraphics[width=1\textwidth]{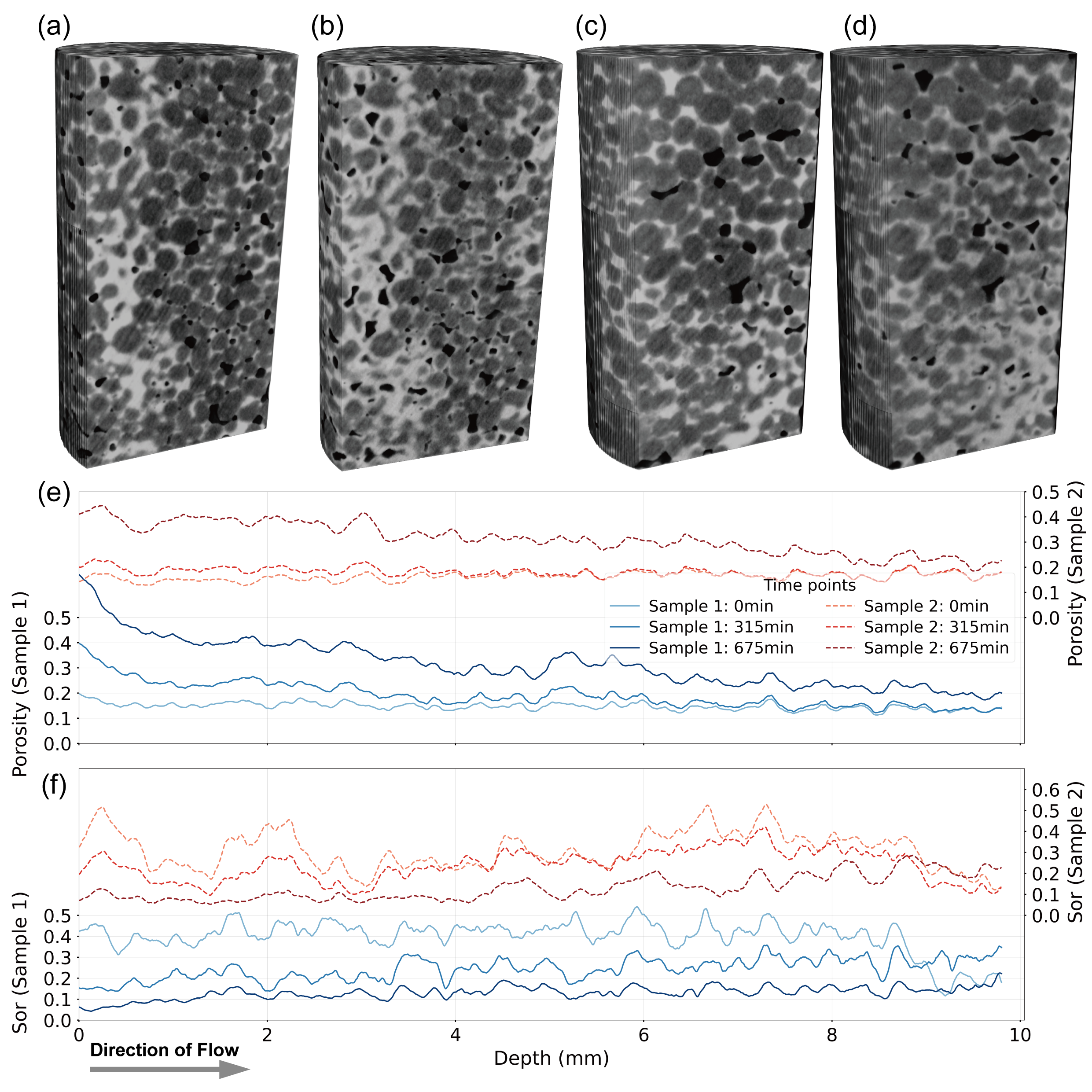}
  \caption{Grayscale images depicting the temporal evolution of dissolution in Sample 1 (a–b) and Sample 2 (c–d) after low flow rate (0.05 ml/min) and after high flow rate (0.5 ml/min). (e) and (f): Porosity and oil saturation profiles for Sample 1 at initial conditions, after low flow rate injection, and after high flow rate injection.}
  \label{fig:7}
\end{figure}

These mechanisms are supported by the flow field evolution (Figure~\ref{fig:6}c-h). At higher injection rates, the velocity distribution narrows, with a marked reduction in stagnant voxels and velocity streamlines become more symmetric. This homogenization of the flow field signifies more uniform reactant delivery and efficient dissolution. This enhanced dissolution is further illustrated in Figure~\ref{fig:7}, which shows the grayscale images of Sample 1 and Sample 2 after low and high flow rates (Figure~\ref{fig:7}a-d), as well as the corresponding porosity and oil saturation profiles (Figure~\ref{fig:7}e-f). The high flow rate promotes a rapid increase in porosity and more effective oil displacement resulting in Sor decrease in both samples. In Sample 1, due to the formation of preferential flow paths, the oil from downstream is more easily transported to the highly conductive channels, forming large oil ganglia. In Sample 2 higher injection rates remobilizes  the oil which was blocking the further channel development 3 mm from the inlet at the low injection rate (recall Figure~\ref{fig:3}l).  This is seen as a decrease in oil saturation throughout the whole sample  in Figure~\ref{fig:3}.

Ultimately this study further demonstrates the existence of the non-uniform dissolution regime (also termed channeling) characterized by moderate Pe, where both diffusion and advection are contributing to transport, and low Da, as encountered in CO$_2$ storage reactive transport. It reveals that the interplay between pore space heterogeneity, oil distribution, displacement and flow rate dictate the dissolution pattern, which in turn governs the overall reaction rate by controlling the efficiency of mass transport.

By offering a detailed pore-scale analysis and perspective on how heterogeneity and the presence of hydrocarbon control flow and reaction dynamics, this study provides valuable insights for optimizing carbon capture and storage (CCS) strategies and predicting the behavior of carbonate reservoirs\cite{spycherCO2H2OMixturesGeological2005} and groundwater remediation \cite{steefelCesiumMigrationHanford2003}, as well as critical global challenges including ocean acidification \cite{tribollet2009effects} and nuclear waste disposal \cite{solerInteractionHyperalkalineFluids2005}.
Moreover, the integration of pore-scale insights into continuum-scale models enhances the accuracy of reactive transport predictions, effectively connecting micro-scale processes to macro-scale implications. This research supports the development of sustainable strategies for carbon management and environmental protection.

\section*{Acknowledgements}

QM gratefully acknowledges Resource Geophysics Academy, Imperial College London for financial support. The authors also extend their sincere gratitude to Anindityo Patmonoaji, Edward Bailey, and Vincenzo Cunsolo for their invaluable assistance in conducting the experiments and analyzing the results.

\newpage
\section*{Appendix A}

\renewcommand{\thefigure}{A\arabic{figure}}
\setcounter{figure}{0}

\noindent\textbf{Péclet-Damköhler Number}
\begin{figure}[H]
  \centering
  \includegraphics[width=0.9\textwidth]{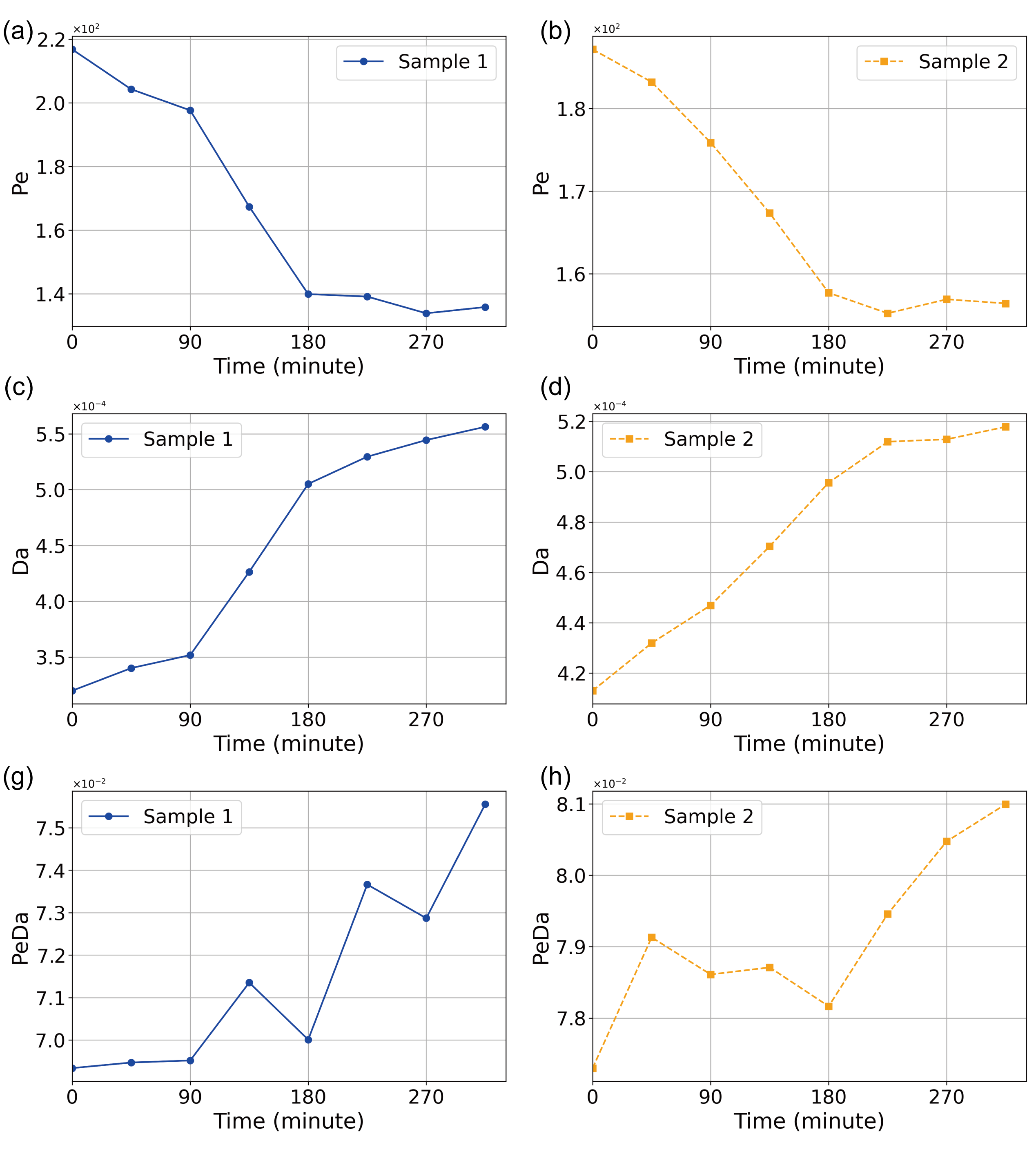}
  \caption{(a)-(f): Dimensionless numbers (Pe, Da, and PeDa) as a function of experimental time for Samples 1 and 2 (0–315 minutes).}
  \label{fig:a1}
\end{figure}

\FloatBarrier   

\bibliography{main}

\end{document}